\documentclass[11pt,preprint2]{aastex}

\def\lesssim{\mathrel{\hbox{\rlap{\hbox{\lower4pt\hbox{$\sim$}}}\hbox{$<$}}}}
\def\gtrsim{\mathrel{\hbox{\rlap{\hbox{\lower4pt\hbox{$\sim$}}}\hbox{$>$}}}}
\newcommand{\mincir}{\raise -2.truept\hbox{\rlap{\hbox{$\sim$}}\raise5.truept
\hbox{$<$}\ }}
\newcommand{\magcir}{\raise -2.truept\hbox{\rlap{\hbox{$\sim$}}\raise5.truept
\hbox{$>$}\ }}
\newcommand{\siml}{\raise -2.truept\hbox{\rlap{\hbox{$\sim$}}\raise5.truept
\hbox{$<$}\ }}
\newcommand{\simg}{\raise -2.truept\hbox{\rlap{\hbox{$\sim$}}\raise5.truept
\hbox{$>$}\ }}
\newcommand{\be}{\begin{equation}}
\newcommand{\ee}{\end{equation}}
\newcommand{\ba}{\begin{eqnarray}}
\newcommand{\ea}{\end{eqnarray}}
\newcommand {\h} {$h^{-1}$ Mpc $ \;$}
\newcommand {\kpc} {$h^{-1}$ kpc}
\newcommand {\hh} {$h^{-1}$ Mpc}
\newcommand {\ks} {km~s$^{-1} \;$}
\newcommand {\kss} {km~s$^{-1}$}
\newcommand {\msun} {$h^{-1} \  M_{\odot} \;$}

\newcommand{\Mn}{$M_{\rm n}$}

\defcitealias{vdm00}{vdM00}
\defcitealias{nfw97}{NFW}
\defcitealias{dia99}{D99}
\defcitealias{dep02}{DP02}
\defcitealias{mer87}{Merritt 1987}
\shorttitle{Cluster Mass Profile}
\shortauthors{Biviano \& Girardi}

\begin{document}


\title{THE MASS PROFILE OF GALAXY CLUSTERS \\ OUT TO $\sim 2 r_{200}$} 


\author{Andrea Biviano}
\affil{INAF -- Osservatorio Astronomico di Trieste, 
Via Tiepolo 11, I-34131 Trieste, Italy}
\email{biviano@ts.astro.it}
\and 
\author{Marisa Girardi}
\affil{Dipartimento di Astronomia, Universit\`{a} 
degli Studi di Trieste, Via Tiepolo 11, I-34131 Trieste, Italy}
\email{girardi@ts.astro.it}


\begin{abstract}
We use the public release of $\simeq 100,000$ galaxies of the {\em Two
Degree Field Galaxy Redshift Survey} (2dFGRS) to analyze the internal
dynamics of galaxy clusters. We select 43 non-interacting clusters
which are adequately sampled in the 2dFGRS public release. Members of
these clusters are selected out to $\sim 2$ virial radii.  We build an
ensemble cluster by stacking together the 43 clusters, after
appropriate scaling of their galaxy velocities and clustercentric
distances. We solve the Jeans equation for the hydrostatic equilibrium
for the member galaxies within the virial radius of the ensemble
cluster, assuming isotropic orbits. We constrain the cluster mass
profile within the virial radius by exploring parameterized models for
the cluster mass-density profile. We find that both cuspy profiles and
profiles with a core are acceptable. In particular, the concentration
parameter of the best fit NFW model is as predicted from numerical
simulations in a $\Lambda$CDM cosmology. Density profiles with very
large core-radii are ruled out.  Beyond the virial radius, dynamical
equilibrium cannot be taken for granted, and the Jeans equation may
not be applicable. In order to extend our dynamical analysis out to
$\sim 2$ virial radii, we rely upon the method which uses the
amplitude of caustics in the space of galaxy clustercentric distances
and velocities. We find very good agreement between the mass profile
determined with the caustic method and the extrapolation to $\sim 2$
virial radii of the best-fit mass profile determined by the Jeans
analysis in the virialized inner region. We determine the
mass-to-number density profile, and find it is fully consistent with a
constant within the virial radius. The mass-to-number density profile
is however inconsistent with a constant when the full radial range
from 0 to $\sim 2$ virial radii is considered, unless the sample used
to determine the number density profile is restricted to the
early-type galaxies.
\end{abstract}


\keywords{ galaxies: clusters: general -- galaxies: kinematics and
dynamics -- dark matter -- cosmology: observations}


\section{INTRODUCTION}\label{intro}

The determination of cluster masses has a long story \citep[see,
e.g.][for a review]{b02}. It dates back to \citet{zwi33,zwi37}'s
and \citet{smi36}'s preliminary estimates of the masses of the Coma
and the Virgo cluster. These early estimates were based on the virial
theorem, using galaxies as unbiased tracers of the cluster potential
(i.e. the {\em light traces mass} hypothesis).  \citet{the86} and
\citet{mer87} were among the first to point out that cluster masses
were in fact poorly known, since relaxing the {\em light traces mass}
assumption widens the range of allowed mass models considerably.
However, the limited amount of redshift data made it very difficult,
if not impossible, to constrain the relative distributions of cluster
mass and cluster light.

With the advent of multi-object spectroscopy, a large number of
redshifts for cluster galaxies became available. In particular, the
two main catalogs of redshifts of cluster galaxies became available,
viz. the ESO Nearby Abell Cluster Survey
\citep[ENACS,][]{kat96,kat98}, and the Canadian Network for
Observational Cosmology \citep[CNOC,][]{yee96,ell98}. Moreover, field
galaxy surveys also contributed to increase the number of redshifts
for cluster galaxies, in particular, the Century Survey \citep{weg01},
and the Two Degree Field Galaxy Redshift Survey
\citep[2dFGRS,][hereafter DP02]{col01,dep02}.

These new data-bases prompted new more detailed investigations in the
issue of the cluster mass determination.  With a significant number of
galaxy redshifts per cluster, it became possible to determine the
cluster mass profile by solving the Jeans equation
\citep[e.g.,][]{bin87} for the equilibrium of member galaxies in the
cluster gravitational potential. Often, several cluster samples have
been combined to improve the statistics, by making the implicit
assumption of cluster homology.

From the analysis of $\sim 1000$ galaxies in the CNOC clusters,
combined to form a single cluster, \citet{car97a} concluded that
galaxies trace the mass to within $\pm 30$\%, and that the cluster
mass profile is well described by a \citet[NFW hereafter]{nfw97}
profile, or by a \citet{her90} profile. Their results have
substantially been confirmed by the more detailed analysis of
\citet[hereafter vdM00]{vdm00}. \citetalias{vdm00} found that the
mass-to-light ratio is almost independent of radius, and the
\citetalias{nfw97} profile is an adequate fit to the data, but other
density profiles are equally acceptable, depending on the orbital
anisotropy of cluster galaxies.  \citet{car01} have recently applied
the technique of \citet{car97a} to a sample of $\sim 700$ galaxies in
$\sim 200$ groups from the CNOC2. They found a strongly increasing
mass-to-light ratio with radius, a result at variance with that found
by \citet{mah99}. A detailed analysis of the mass and anisotropy
profiles of clusters from ENACS is ongoing (Katgert, Biviano \&
Mazure, in preparation).

A new technique for the determination of cluster mass profiles has
recently been introduced by \citet{dia97} and \citet[hereafter
D99]{dia99}. It is based on the determination of the amplitude of the
caustics in the space of velocities and clustercentric
radii. Applications of this technique include \citet*{gel99}'s and
\citet{rin01}'s determination of the Coma cluster mass profile out to
14 \hh, \citet{rin00}'s determination of the mass profile of Abell~576
out to 6 \hh, and \citet{rei00}'s determination of the mass of the
Shapley supercluster. The \citetalias{nfw97} profile was found to
provide a consistent fit to the mass profiles. Interestingly,
\citet{rin00} found a decreasing mass-to-light profile in Abell~576
while \citet{rin01} found a constant mass-to-light profile in Coma.

From a theoretical point of view, both decreasing and increasing
mass-to-light profiles have been predicted. A decreasing mass-to-light
profile could result from the tidal stripping of galaxy halos in the
cluster centers \citep{mam00}. An increasing mass-to-light profile
could instead result from the combined effects of dynamical friction
and galaxy merging \citep{fus98}. In general, galaxies are unlikely to
be distributed exactly like the mass, if anything because different galaxy
populations have different distributions \citep[e.g.,][]{biv02}.

In this paper we determine the mass profile of an ensemble cluster
built from the combined data of 43 clusters extracted from the 2dFGRS
\citep{col01}, using the techniques described by \citetalias{vdm00}
and \citet[see also \citetalias{dia99}]{dia97}. Since the 2dFGRS is a
field survey, only a small fraction of the 2dFGRS galaxies are cluster
members. Our final sample contains 1345 cluster members only, a number
comparable to that of cluster members in the CNOC and smaller than
that of cluster members in the ENACS. The main advantage of using the
2dFGRS for the determination of the cluster mass profile is the
possibility of sampling the cluster dynamics to a large distance from
the cluster center.

This paper is organized as follows.  In \S~\ref{data} we describe the
data sample, how we assign the cluster membership and how we combine
the 43 cluster data-sets into a single ensemble cluster. In
\S~\ref{mprof} we determine the ensemble cluster mass profile. In
\S~\ref{mton} we use this mass profile to determine the mass-to-number
density profile. In \S~\ref{disc} we discuss our results and compare
them with previous determinations of cluster mass and mass-to-light
profiles.  Finally, in \S~\ref{summ} we present a summary of our
results. A Hubble constant of 100 $h$ \ks Mpc$^{-1}$ is used
throughout.

\section{THE DATA SAMPLE}\label{data}

The data sample we use is the 2dFGRS public release version of June,
30$^{th}$, 2001. It contains $\sim 100,000$ redshifts and spectral
types as well as $b_{\mbox{J}}$ photometry for $\sim 460,000$ objects.
The $b_{\mbox{J}}$ magnitudes are extinction-corrected total
magnitudes derived from updated versions of the original APM
(Automated Plate Measuring machine) scans
\citep[see][]{col01,nor01}. The target galaxies for the 2dFGRS were
selected to have $b_{\mbox{J}} \leq 19.45$.  Within the 2dFGRS public
release, we only consider those galaxies in the fields of galaxy
clusters. Galaxy clusters have been identified in the 2dFGRS by
\citetalias{dep02}, via a cross-correlation of the 2dFGRS with the
cluster catalogs of \citet[hereafter, ACO]{aco89}, of \citet[the
Edinburgh-Durham Cluster Catalogue, EDCC]{lum92} and of \citet[the APM
cluster catalog]{dal97}. From \citetalias{dep02}'s sample we exclude
those clusters whose central regions are not evenly sampled in the
2dFGRS public release, and those which, according to the authors
themselves, have less than 20 members with redshifts in the 2dFGRS. Of
course, we also get rid of double entries in \citetalias{dep02}'s list
(such as, e.g., APM~309 which corresponds to ACO~3062). We are thus
left with 91 clusters.

Following \citet{car97a} we define $r_u \equiv \sqrt{3} \,
\sigma_p/[10H(z)] \simeq r_{200}$, the radius at which the mean
interior overdensity is 200 times the critical density (in a
$\Omega_0=1$ Universe), usually called the 'virial radius'. As a first
estimate of the clusters projected velocity dispersions along the
line-of-sight, $\sigma_{p}$'s, we use \citetalias{dep02}'s values.

To these 91 clusters, we apply the selection procedure of cluster
members of \citet{fad96} and \citet{gir98}.  First, we apply
\citet{pis93}'s one-dimensional algorithm based on the adaptive kernel
technique (see also Appendix A of \citealt{gir96}).  The adaptive
kernel technique is a nonparametric method for the evaluation of the
density probability function underlying an observational discrete data
set.  The method returns a list of all density peaks detected, as well
as their probabilities, and the objects associated to each peak. We
apply this method to the velocity distribution of all galaxies
projected within $2 r_u$ of each cluster center.  We select as
significant those peaks with $\geq 98$\% probability.  When
adjacent peaks have 20\% or more of their objects in common, we join
them together if they are closer in velocity than 1500 \kss.  When one
peak is closer than 1500 \ks to two other peaks, themselves separated
by more than 1500 \kss, we join the central peak to the one which is
closest, and leave the third peak separate.

Quite often, several significant density peaks are found in a single
cluster region. Among the significant peaks, we choose that one which
is closest to the mean cluster velocity, as given by \citetalias{dep02}.
In this way we avoid possible mis-identifications, since sometimes the
search region around each cluster (a circle of $2 r_u$ radius)
is wide enough to contain another cluster.

We identify 75 clusters with a significant density peak in the
velocity space and at least 20 galaxies associated to that peak. In
one case (ACO-S1155), no significant peak is found, and in another 15
cases, there are less than 20 galaxies in the peak corresponding to
the cluster. These 16 clusters are excluded from our sample.

We re-define the centers of the 75 clusters.  We define the new
cluster center as the position of the highest density peak found
with a two-dimensional adaptive kernel technique. In order to avoid
mis-identifications, we only use galaxies less distant than 0.5 \h
from the old cluster center listed by \citetalias{dep02}.  Using these
new center determinations, we apply the 'shifting-gap' method
\citep{fad96} to get rid of the remaining interlopers.  The
shifting-gap method makes use of both galaxy velocities and
clustercentric distances.  Galaxies are sorted in order of increasing
distance from their cluster centers.  Among samples of galaxies in
(overlapping and shifting) bins of 0.4 \h from the cluster center (or
wide enough to contain $\geq 15$ galaxies each), gaps $\geq 1000$ \ks
are identified in the galaxy velocity distribution.  These gaps define
the edges of the galaxy velocity distribution. Outliers from this
distribution are flagged as interlopers.

Having excluded the interlopers, we make a new estimate of the cluster
mean velocities, $\overline{v}$'s, and velocity dispersions,
$\sigma_p$'s, (and hence $r_u$'s).  The velocity moments are estimated
with the robust methods of \citet*[see also
\citealt{gir98}]{bee90}. Velocity dispersions are corrected for the
velocity errors and moved to the cluster rest-frames
\citep{dan80}. These new $\overline{v}$, $\sigma_p$ and $r_u$
estimates are the starting values for an iterative procedure. We
consider only the cluster members within a distance $\leq r_u$ from
the cluster center to recalculate $\overline{v}$ and $\sigma_p$. If
the difference between the new estimate of $\sigma_p$ and the previous
one is $\leq 10$\%, the procedure is halted. Otherwise, we iterate, by
recalculating $\overline{v}$ and $\sigma_p$ on the members within a
distance of $\leq r_u$ from the cluster center, where the new value of
$r_u$ is used. With this iterative procedure, we eliminate 7 of the
initial 75 clusters, because they are left with less than 5 members
within $r_u$ at any of the iteration steps.

With the new center and $r_u$ determinations, we then iterate the
adaptive-kernel procedure and the shifting-gap method for the
selection of cluster members.  This time, we extend the search region
to $3 r_u$, in order to ensure that we will be covering a region of at
least $2 r_u$ radius when the final definitive estimate of $r_u$ is
done.  On the selected galaxies we re-determine $\overline{v}$ and
$\sigma_p$ via the iterative procedure described above. We are left
with a sample of 67 clusters, after eliminating one because it
contains less than 5 members within $r_u$.

In order to determine the mass profile of galaxy clusters to
sufficiently large distances, it is wise to exclude interacting
clusters, whose dynamics is difficult to model. Say $\overline{v_i}$,
$\sigma_i$, and $r_{u,i}$ are, respectively, the mean velocity, velocity
dispersion, and virial radius of cluster $i$, and $R_{i,j}$ is the
projected distance between the centers of clusters $i$ and $j$; we
say cluster $i$ and cluster $j$ are 'interacting' if the following
two conditions apply:
\begin{equation}
\mid \overline{v_i}-\overline{v_j} \mid < 3 (\sigma_i + \sigma_j) \\
\label{cond1}
\end{equation}
\begin{equation}
R_{i,j} < 2 (r_{u,i} + r_{u,j}).
\label{cond2}
\end{equation}
We eliminate 24 interacting clusters, and we are thus left with a
sample of 43 reasonably isolated clusters.

Before we can construct azimuthally averaged profiles, we must verify
that the circular region of $2 r_u$ radius around each cluster center
is fully covered in the 2dFGRS public release data sample.  This is
indeed the case for all the clusters in our sample, except two,
APM~715 and ACO~892. For these two clusters we define an angular
sector which excludes the region not covered in the 2dFGRS public
release data sample, and we consider only the galaxies within this
sector.

\begin{figure}
\plotone{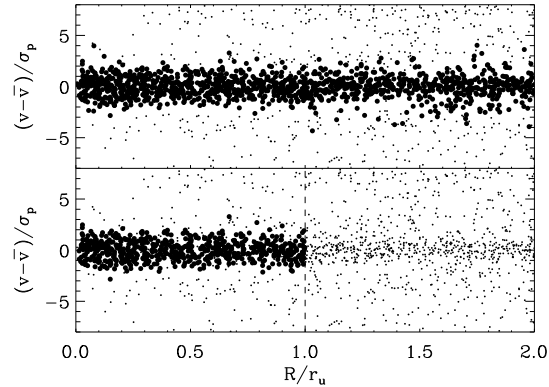}
\caption{Distribution of galaxies of the ensemble cluster
in the space of normalized clustercentric distances
and velocities. Top panel: large dots indicate cluster members.
Bottom panel: large dots indicate the 642
cluster members, late-type spirals excluded,
within $R \leq r_u$, on which the Jeans analysis
is based. The vertical dashed line indicates the $R=r_u$ limit beyond
which the cluster membership assignment is less reliable.
\label{f-rvj}}
\end{figure}

Using a two-dimensional Kolmogorov-Smirnov test \citep[e.g.][]{fasfra}
we also verify that the spatial distributions of galaxies with and
without redshifts are not significantly different, at least out to $2
r_u$.  There are only four clusters for which there is a marginal
($\geq 90$\%) evidence of a difference: ACO~957, ACO~1750, ACO~2734,
EDCC~153. For these clusters the redshift completeness is not
homogeneous within $2 r_u$, and decreases outside. Only 9\%
(respectively 2\%) of all the galaxies in our sample occupy regions
where the completeness in redshift is significantly below
(respectively 50\%) the average completeness for their cluster. We
therefore deem it unnecessary to correct for this incompleteness.
In summary, thanks to the quality of the 2dFGRS, our final sample is
almost unaffected by problems of radial incompleteness.

Our final sample contains 3602 field galaxies and 1345 cluster members
with redshifts, within $2 r_u$ from the centers of 43 clusters. The
absolute $b_{\mbox{J}}$ magnitudes of the 1345 cluster members,
k-corrected as described by \citet{mad02}, range from $M_{b_J}=-22.3+5
\log h$ to $M_{b_J}=-14.4+5 \log h$.  The different clusters are
sampled down to different absolute magnitude limits, since they lie at
different redshifts, from a limiting $M_{b_J}=-14.4+5 \log h$ in
ACO~S301, to a limiting $M_{b_J}=-19.3+5 \log h$ in APM~294.

The 43 clusters have an average $\sigma_p$ of 490 \kss.  This is
significantly lower than the average $\sigma_p$ estimated by
\citetalias{dep02} for the same clusters, 776 \kss.  Interestingly,
our velocity dispersion estimates and those of \citetalias{dep02} are
significantly correlated (97\% probability according to a Spearman
rank correlation test), which suggests that the difference in
$\sigma_p$ is systematic. 

We believe that \citetalias{dep02}'s $\sigma_p$'s are systematically
overestimated, for the following reasons. First, \citetalias{dep02}'s
highest value of $\sigma_p$ among the 43 clusters is 1620 \kss, higher
than any of the $\sigma_p$ values found in the complete cluster sample
of \citet{maz96}. For comparison, the hottest X-ray cluster known,
1E0657-56, has a velocity dispersion of $1201^{+100}_{-92}$ \kss
\citep{barrena}.  Second, \citetalias{dep02}'s $\sigma_p$-estimates
are too large when compared to the cluster richnesses. This can be
seen by considering the 27 ACO clusters in our sample, whose average
richness count is 41. The 27 ACO clusters have an average $\sigma_p$
of 493 \kss according to our analysis, or 822 \kss according to
\citetalias{dep02}.  The former value is as expected on the basis of
the cluster richness-velocity dispersion correlation \citep{fad96},
while the latter value is certainly too high. Finally, our
$\sigma_p$-estimates are in better agreement than those of
\citetalias{dep02}, with the values given in the literature for five
of the 43 clusters \citep{maz96,gir98,alo99}.  The mean difference
between our $\sigma_p$-estimates and those of the literature is $0 \pm
172$ \kss, that between \citetalias{dep02}'s $\sigma_p$-estimates and
those of the literature is $+378 \pm 101$ \kss.

Our cluster sample is presented in Table~\ref{tabsample}. We list
in column (1) the cluster name (as given in the list of \citetalias{dep02}),
in columns (2) and (3) the coordinates of the cluster center,
in column (4) the number of member galaxies within $2 r_u$,
in column (5) the number of member galaxies within $r_u$,
in column (6) the mean cluster velocity, and in column (7) the
cluster velocity dispersion (computed using the robust estimators
of \citealt{bee90}).

\section{THE MASS PROFILE}\label{mprof}

In order to put significant constraints on the cluster mass profile we
need a sufficiently large data-set. We therefore join together the 43
clusters into a single ensemble cluster. To this aim, we scale the
projected distances of galaxies from their cluster centers, $R$'s, by
$r_u$ (see \S~\ref{data}) which is an approximate estimate of
$r_{200}$. We scale the line-of-sight velocities of the galaxies
relative to the cluster mean velocity, $(v-\overline{v})$, with the
global velocity dispersions $\sigma_p$ of their parent
clusters. Similar scalings have often been used in the literature
\citep[e.g.][]{biv92,car97a}. 

\begin{figure}
\plotone{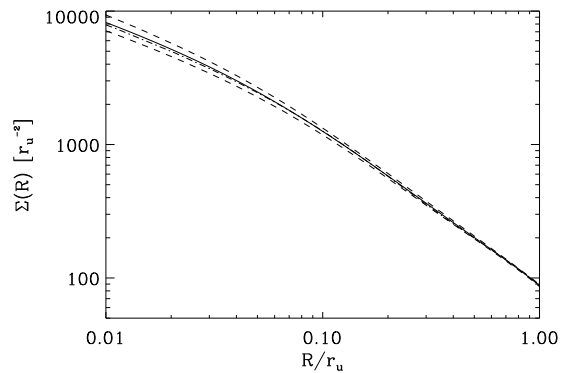}
\caption{Projected number density profile of the cluster members
(late-type spirals excluded), determined through the MAPELN
method. Error bars were determined from 1000 bootstrap resamplings
(dashed lines). The dash-dotted line represents the best-fit to the
observed profile with a function of the form suggested by
\citetalias{vdm00}. \label{f-ir}}
\end{figure}

In Figure~\ref{f-rvj} (top panel) we show the distribution of galaxies
of the ensemble cluster in the space of normalized clustercentric
distances and velocities.  Large dots indicate cluster members. It is
easy to spot a few apparent outliers that have been classified as
cluster members, mostly at large distances from the cluster
center. These possible misidentifications are due to the difficulty of
the shifting-gap procedure in rejecting interlopers in regions of low
galaxy density. In order to avoid possible problems of wrong
membership assignments, we do not use the membership assignment beyond
$r_u$ in our analysis.

\subsection{The Jeans approach}\label{jeans}
The determination of the mass profile can in principle be obtained by
a straightforward application of the Jeans equation. A reasonable
assumption must however be made on the orbital anisotropy of the
tracers of the gravitational potential. Several analyses have reached
the conclusion that the orbits of galaxies in clusters are
quasi-isotropic \citep{car97b,vdm00}, except for late-type spirals, which
are found to be on moderately radial orbits \citep[see, e.g.,][see also
\citealt*{ada98}]{moh96,biv97}. By excluding late-type spirals from our
sample we are therefore confident that we can assume isotropic orbits
for the remaining cluster galaxies. In order to exclude late-type spirals
we impose an upper limit on the spectral type parameter provided by
the 2dFGRS, $\eta < 1$ \citep[see][]{mad02}.

As explained in \S~\ref{mprof}, our membership assignment may not be
very robust beyond $r_u$. We therefore restrict the Jeans analysis to
the sample of cluster members with $R \leq r_u$. Additional
motivations for this choice are:
\begin{itemize}
\item reducing the effect of possible anisotropies.  In fact, galaxies
infalling into a cluster are likely to lose their radial anisotropy
when crossing the denser central regions \citep{mam95}. 
\item Reducing
the influence of possible subclustering. In fact, the central cluster
regions are a hostile environment for the survival of substructures
\citep{gon94}. 
\item Satisfying the condition of dynamical equilibrium,
needed for the application of the Jeans equation. In fact, $r_u
\approx r_{200}$ delimits the region of virialization.
\end{itemize}

In total, we have 642 cluster members with $R \leq r_u$ and $\eta <
1$. Their distribution in the space of clustercentric distances and
velocities is shown in Figure~\ref{f-rvj} (bottom panel).  Their
normalized velocity-distribution is not significantly different from a
Gaussian, according to a Kolmogorov-Smirnov test. This suggests that
our assumption of isotropy is acceptable
\citepalias[see][]{mer87,vdm00}. We solve the Jeans equation under the
isotropic assumption for the sample of 642 cluster members. For this,
we follow the method of \citet{vdm94} and \citetalias{vdm00}.

First, we determine the projected number density profile, $\Sigma(R)$ of
cluster members out to $2 r_u$ (late-type spirals excluded) with the
Maximum Penalized Likelihood (MAPELN) method of \citet{mer94}. We fit
the observed profile with a multi-parameter function of the form
suggested by \citetalias{vdm00} (see eq.[2] in \citetalias{vdm00}, and
Figure~\ref{f-ir}).  The function is sufficiently general as to allow
a very close representation of the real $\Sigma(R)$.  We Abel-invert
the $\Sigma(R)$ best-fitting function to produce the 3-dimensional
number density profile $\nu(r)$. Abel-inversion requires
the knowledge of the asymptotic behaviour of $\Sigma(R)$ to large
radii. We try different extrapolations of the observed $\Sigma(R)$
to $10 r_u$ and choose the one that gives the $\nu(r)$ which most
closely match the obsered $\Sigma(R)$ after Abel-projection.

We then consider two functional forms for the mass density profile. 
One is taken from \citetalias{vdm00}:
\begin{equation}
\rho(r)=\rho_0 (r/a)^{-\xi} (1+r/a)^{\xi-3}. \label{eq1}
\end{equation}
The other, often employed in the literature \citep[e.g.][]{gir98},
is the $\beta$-model \citep{cav78} density profile:
\begin{equation}
\rho(r)=\rho_0 [1+(r/r_c)^2]^{-3 \beta/2}. \label{eq2}
\end{equation}
Hereafter we refer to these two mass-density models as the $\xi$- and the
$\beta$-model.

We consider a grid of values for the mass density parameters $(a,\xi)$
and $(r_c,\beta)$. For each set of values we compute the shape of the
predicted $\sigma_p$ and the density normalization $\rho_0$ that
minimizes the difference between the predicted $\sigma_p$ and the
observed (binned) $\sigma_p$, out to $R=r_u$. We then constrain the
allowed values of the density profile parameters in a $\chi^2$ sense.
The observed $\sigma_p$-profile is determined with the biweight
estimator \citep{bee90} in 11 radial bins, each containing 57
galaxies.

\begin{figure}
\plotone{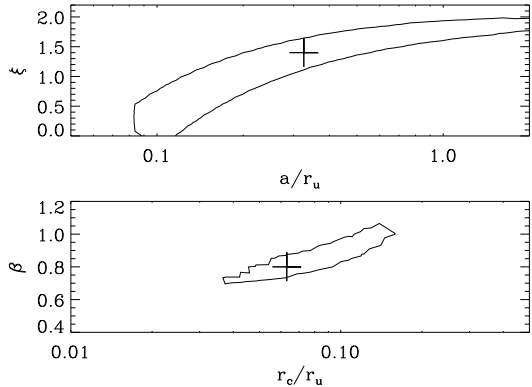}
\caption{Isocontours of allowed parameter values for the mass density
$\xi$ and $\beta$ models -- see text. Crosses indicate
the best-fit values. Top panel: $\chi^2$ 68\% c.l. isocontours in
the space of $(a,\xi)$ parameters. Bottom panel: $\chi^2$ 68\%
c.l. isocontours in the space of $(r_c,\beta)$ parameters.
\label{f-chi2}}
\end{figure}

\begin{figure}
\plotone{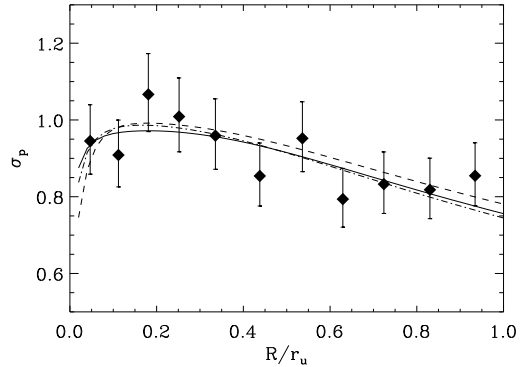}
\caption{Predicted velocity dispersion profiles from the best-fit mass
density models, and the observed $\sigma_p$ (binned points with 68\%
error bars). Solid line: $\xi$-model ($a=0.33 \, r_u$ and $\xi=1.4$);
dash-dotted line: \citetalias{nfw97}-model ($a=0.18 \, r_u$ and $\xi=1$);
dashed line: $\beta$-model ($r_c=0.06 \, r_u$ and $\beta=0.8$).
\label{f-vdpfit}}
\end{figure}

\begin{figure}
\plotone{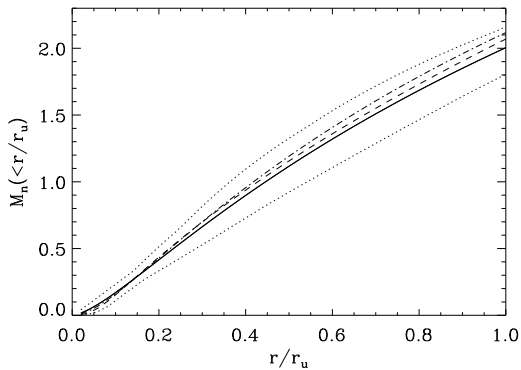}
\caption{Mass profiles determined by the Jeans analysis. Solid line:
mass profile obtained from the best-fit $\xi$ density model;
dash-dotted line: mass profile obtained from the best-fit
\citetalias{nfw97} density model; dashed line: mass profile obtained
from the best-fit $\beta$-model. Dotted lines: 68\% c.l. for the mass
profiles, derived from integration of the density profiles which
provide an acceptable fit (to within the 68\% c.l.) to the observed
$\sigma_p$ (see Figure~\ref{f-chi2}).
\label{f-massj}}
\end{figure}

The allowed 68\% confidence level (c.l. hereafter) isocontours in the
space of $(a,\xi)$, and $(r_c,\beta)$ parameters are shown in
Figure~\ref{f-chi2}. The confidence levels are determined as described
in \citet{avn76}, for two interesting parameters. Clearly, density
models with a wide range of parameters provide an acceptable fit to
our data. In particular, the best fit to the $\xi$-model is obtained
for $a=0.33 \, r_u$ and $\xi=1.4$, with a reduced
$\chi^2=0.6$. A standard \citetalias{nfw97} model (corresponding to
the case $\xi=1.0$) provides an almost equally good fit to the
data. The best fit \citetalias{nfw97} model is obtained for $a=0.18 \, r_u$,
as expected for halos with $\sigma_p \simeq 500$ \ks in a
$\Lambda$CDM Universe. Also acceptable are the density
models with a core ($\xi=0$), in which case the scale is constrained
to be small, $a \simeq 0.1 \, r_u$. Consistently, the allowed values of
$r_c$ for the $\beta$-model are also rather small.  The best-fit is
obtained for $r_c=0.06 \, r_u$ and $\beta=0.8$, which are close to the values
found by \citet{gir98} for the number density profiles of cluster
galaxies. Core-radii $r_c \geq 0.2 \, r_u$ ($r_c \geq 0.3 \, r_u$) are
excluded at the 68\% (90\%) c.l.. A standard \citet{kin62} profile
($\beta=1.0$) with $r_c=0.14 \, r_u$ is also an acceptable fit; these values
are close to those obtained by \citet{ada98} for the number density
profiles of cluster galaxies.

In Figure~\ref{f-vdpfit} we show the predicted velocity dispersion
profiles for the best-fit $\xi$- (solid line), $\xi=1$
\citetalias{nfw97} (dash-dotted line), and $\beta$- (dashed-line)
models, as well as the observed (binned) $\sigma_p$-profile.

We obtain the 68\% c.l. for the cluster mass profile by integrating
the mass density profiles which provide 68\% c.l. acceptable fits to
the observed $\sigma_p$-profile.  The mass profiles determined via
integration of the best-fit mass density profiles, and the 68\%
c.l. for the cluster mass profile are shown in
Figure~\ref{f-massj}. The mass is in the normalized units of our
ensemble cluster, i.e. $r_u \sigma_p^2/G$, which corresponds to
$\approx 5 \times 10^{13}$ \msun, for the average value of $\sigma_p
\simeq 500$ \ks of the 43 clusters that constitute our ensemble
cluster. We call \Mn~ the normalized mass. The mass profiles obtained
from the best-fit $\xi$- and $\beta$- mass density models are very
similar in the fitted region ($r \leq r_u$). The total mass at $r \leq
r_u$ appears to be determined with an accuracy of $\simeq \pm 8$\%.

\subsection{The caustic approach}\label{caustic}

\begin{figure}
\plotone{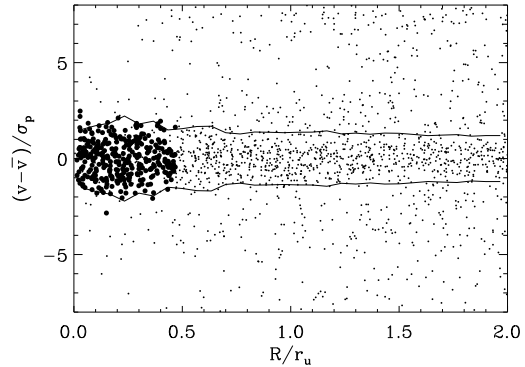}
\caption{'Optimal caustic' and the distribution of galaxies from the
ensemble cluster in the space of normalized clustercentric distances
and velocities. Large dots indicate the
cluster members used to determine $\overline{v_{\rm n}^2}(\leq
\overline{R})$ -- see text.
\label{f-caustic}}
\end{figure}

In order to extend the determination of the cluster mass profile
beyond the virialization radius $r_u \simeq r_{200}$, we must rely on
methods that do not assume the cluster dynamical equilibrium, since
this is unlikely to hold in the external cluster regions.  The method
we use is that of \citet{dia97} and \citetalias{dia99}.
These authors have shown that the velocity field in the cluster
outskirts is determined by the cluster mass distribution, via
\begin{equation}
GM(<r)-GM(<r_0)=\int_{r_0}^r {\cal A}^2(x) \, {\cal F}_{\beta}(x) \, dx,
\label{eqdiaf}
\end{equation}
where ${\cal A}$ is the amplitude of the caustics described by the
galaxy distribution in the space of normalized clustercentric
distances and velocities, and ${\cal F}_{\beta}$ is a function of the
gravitational potential and the anisotropy profile.

We determine the caustics with an adaptive kernel method, as described
in \citetalias{dia99} (see also \citealt{pis96}).  Following
\citet{dia97} and \citet{gel99}, we scale the smoothing-window sizes
along the axes of clustercentric distance and velocities before
applying the adaptive kernel method, in such a way as to give equal
weights to the typical uncertainties of normalized clustercentric
distances and velocities. In our sample the median velocity
uncertainty is 90 km~s$^{-1}$, or 0.2 in normalized units.  The
positional uncertainty of a galaxy is of the order of the galaxy size,
i.e.  about 20 \kpc, or 0.02 in normalized units.  We therefore take
$q=10$ as the ratio between the smoothing-window sizes on the two
axes. \citet{gel99} have shown that the mass profile determination is
largely independent on the precise choice of $q$. As in \citet{gel99},
we symmetrize the caustics with respect to the mean cluster velocity.

\begin{figure}
\plotone{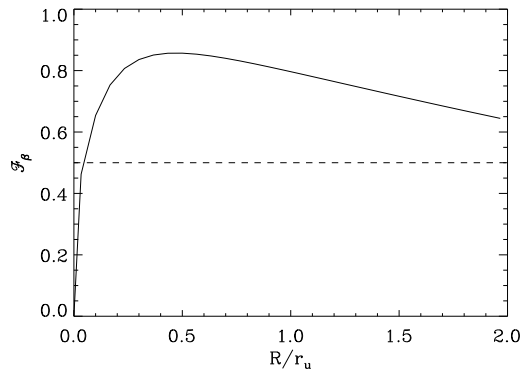}
\caption{Function ${\cal F}_{\beta}$ for the determination of the
cluster mass profile with the caustic method. The dashed line shows the value
${\cal F}_{\beta}=0.5$ used by \citet{gel99}.
\label{f-fbeta}}
\end{figure}

Following \citetalias{dia99} and \citet{gel99}, we choose the caustic
that minimizes the function:
\begin{equation}
S \equiv \mid \overline{v_{esc}^2} - 4 
\overline{v_{\rm n}^2}(\leq \overline{R}) \mid^2. \label{eq4}
\end{equation}
The $\overline{v_{esc}^2}$ term is determined from the caustic
amplitude, and $\overline{v_{\rm n}^2} (\leq \overline{R})$ is
the normalized velocity dispersion of the cluster members contained within
$\overline{R}$, the average clustercentric distance of all
cluster members. As described in \S~\ref{jeans} our fiducial sample of
cluster members does not include late-type spirals and is restricted to
galaxies within $r_u$. We show in Figure~\ref{f-caustic} the caustic
that minimizes the $S$ function (see eq.~\ref{eq4}). Hereafter, we
refer to this caustic as the 'optimal caustic'.

\begin{figure}
\plotone{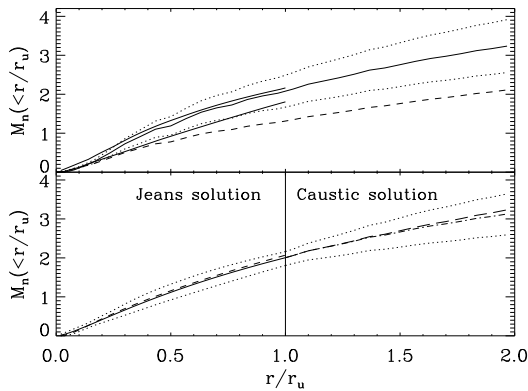}
\caption{Mass profiles determined by the caustic method. Top panel:
the mass profile determined using the ${\cal F}_{\beta}$ function
(solid line) within its 68\% c.l. (dotted lines).  The dashed line
shows the mass profile determined using ${\cal F}_{\beta}=0.5$. Thin
solid lines truncated at $r=r_u$ indicate the 68\% c.l.'s on the mass
profile determined by the Jeans analysis (see \S~\ref{jeans}).  Bottom
panel: the combined mass profile determined via the Jeans and the
caustic methods. The two profiles at $r \leq r_u$ are obtained by
integration of the best-fit $\xi$- and $\beta$- density models (solid and
dashed lines, respectively).  The two profiles at $r > r_u$ are
obtained via the caustic method, by imposing continuity with the $\xi$
mass profile at $r=r_u$, assuming either a variable or a constant
${\cal F}_{\beta}$ (dash-dotted and long-dashed lines, respectively).
Dotted lines indicate the 68\% c.l.'s on the mass profile determined
via the Jeans analysis at $r \leq r_u$, and via the caustic analysis
at $r > r_u$ (continuity is imposed at $r=r_u$).
\label{f-mass}}
\end{figure}

In order to translate the optimal caustic amplitude into a mass
profile, we must know the function ${\cal F}_{\beta}$. This function
actually depends on the gravitational potential one is willing to
determine. However, numerical simulations show that this dependence is
small at sufficiently large distances from the cluster center
\citepalias{dia99}. \citet{dia97} and \citet{gel99} used the simple
approximation ${\cal F}_{\beta} \equiv 0.5$, but this is not really
supported by the results of \citetalias{dia99} (see Figure~3 in
\citetalias{dia99}'s paper). In our analysis we adopt a non-constant
${\cal F}_{\beta}$ (see Figure~\ref{f-fbeta}) which is a smooth
approximation of the result obtained by \citetalias{dia99}
for a $\Lambda$CDM cosmology. The resulting mass function is shown in
Figure~\ref{f-mass} (top panel). The c.l.'s on this mass function are
determined as in \citet{gel99}.  For the sake of comparison, we also
show the mass function one would obtain by adopting ${\cal
F}_{\beta}=0.5$.

As can be seen from the top panel of Figure~\ref{f-mass}, the mass
profiles determined from the same optimal caustic, using different
${\cal F}_{\beta}$ functions, can differ considerably. As a matter of
fact, the choice of ${\cal F}_{\beta}$ constitutes the dominant source
of uncertainty in this method. Note however that the mass profile
determined using a non-constant ${\cal F}_{\beta}$ is in remarkable
consistency with the mass profile determined via the Jeans analysis
(see \S~\ref{jeans}).

In order to reduce the systematic errors in the mass profile, it is
better to avoid using the caustic method in the central cluster
regions. In fact, the main advantage of the caustic method lies in its
applicability beyond the virialized central region, where classical
methods fail, but classical methods may prove superior in the central
region, where the caustic method suffers from the uncertainty in the
function ${\cal F}_{\beta}$.  We can therefore take advantage of the
constraints imposed by the Jeans analysis on the mass interior to
$r_u$, and compute the mass from the caustic method only at $r
> r_u$ (i.e. we set $r_0=r_u$ in eq.~\ref{eqdiaf}). By doing this we
reduce the uncertainty in the mass profile determined via the caustic
method, because the radial dependence of ${\cal F}_{\beta}$ is much
smaller at large radii than in the central region (see Figure~2 in
\citetalias{dia99}). The combined mass profile from the Jeans and the
caustic approach is shown in the bottom panel of
Figure~\ref{f-mass}. Clearly, the mass profiles determined with
different choices of ${\cal F}_{\beta}$ are now in good agreement,
since we compute them only for $r > r_u$.  The uncertainty on
the total mass at $r=2 r_u$ is $\approx \pm 15$\%.

We note that the mass profile determined by the caustic method follows
very closely the extrapolation of the best-fit $\xi$-profile out to
$r=2 r_u$. The best-fit $\beta$-profile is somewhat steeper, but well
within the 68\% confidence region of the caustic mass profile.

\section{THE MASS-TO-NUMBER DENSITY PROFILE}\label{mton}

In \S~\ref{caustic} we have determined the average cluster mass
profile out to $r=2 r_u$. We obtain the average cluster mass density
profile, $\rho$, by numerical differentiation of the mass profile
shown in the bottom panel of Figure~\ref{f-mass} (indicated by the solid
line up to $r=r_u$, and by the dash-dotted line at larger radii). We
determine the 68\% confidence interval on this density profile by
constructing a set of density profiles which are built
non-parametrically from the differentiation of several mass profiles
spanning the allowed 68\% confidence region. In building these density
profiles we impose the physical condition of a negative logarithmic
derivative of the mass density at all radii.

In order to determine the 3-dimensional number density profile, $\nu$
(and its confidence intervals), up $r=2 r_u$, we use MAPELN.  In this
case we do not exclude late-type spiral members, since we are
interested in the total number of galaxies in the cluster,
irrespective of their kinematics. Since the 2dFGRS is complete in {\em
apparent} magnitude, and our 43 clusters span a significant redshift
range (0.02--0.13), we need to impose an {\em absolute} magnitude
limit to ensure the same sampling of the luminosity functions of all
our 43 clusters. We therefore select the 399 cluster members brighter
than the k-corrected \citep{mad02} absolute magnitude $M_{b_J}=-19.3+5
\log h$, which is the magnitude of the faintest member of the most
distant among our 43 clusters, APM~294 (see \S~\ref{data}). This
magnitude is sufficiently bright as to avoid possible problems related
to the increase in the number of dwarf galaxies with clustercentric
distance \citep[see, e.g.][]{bei02,dur02,lob97}.

Since the cluster membership assignment is problematic beyond $r_u$,
we correct the MAPELN density profile at larger distances, by making
use of the optimal caustic. Specifically, beyond $r_u$ we multiply the
MAPELN density profile by the fraction of cluster members which are
contained within the optimal caustic in each given radial bin. This
fraction is 0.8 on average, i.e. $\sim 20$\% of our (presumed) cluster
members beyond $r_u$ could be interlopers.

\begin{figure}
\plotone{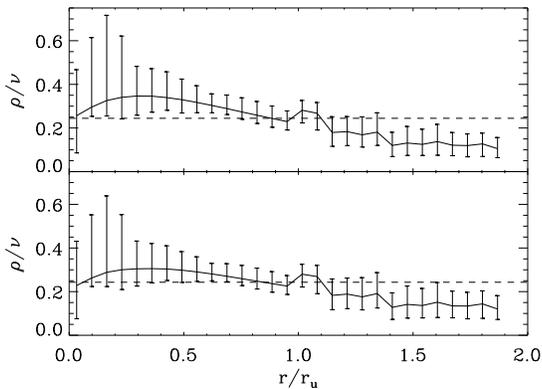}
\caption{Mass-to-number density profiles for the ensemble cluster
(solid lines) and best-fit constants (dashed lines). Top panel: the
number density profile is determined for the sample of all cluster
members brighter than $M_{b_J}=-19.3+5 \log h$. Bottom panel: the
number density profile is determined for the sample of early-type
cluster members brighter than $M_{b_J}=-19.3+5 \log h$.  The error
bars account for the uncertainties in both the mass and the number
density profiles. The vertical scale is in normalized units (see
\S~\ref{jeans}).
\label{f-dmdn}}
\end{figure}

The mass-to-number density ratio as a function of the clustercentric
radial distance is shown in Figure~\ref{f-dmdn} (upper panel).  The
vertical scale in this plot is in normalized units (see
\S~\ref{jeans}).  The irregular behavior at radii larger than $r_u$ is
due to the use of a non-parametric mass profile estimator (the caustic
method).  The bump at $r \simeq r_u$ is related to the fact that we
have joined the mass profile determined from the Jeans analysis and
the mass profile determined from the caustic method, without imposing
a continuous derivative. Apart from obvious irregularities, one can
notice an almost monotonous decrease from $r \simeq 0.3 \, r_u$ to the
outermost radius. We test the null hypothesis of a constant
mass-to-number density profile by the $\chi^2$ method. The null
hypothesis is fully consistent with the data (3.7\% c.l.) when only
the inner $r \leq r_u$ region is considered, but it is rejected
(99.9\% c.l.) over the full radial range ($0<r<2 r_u$). Had we not
corrected for the presence of doubtful cluster members at radii larger
than $r_u$, the decrease would have been (marginally) even stronger.

In Figure~\ref{f-dmdn} (lower panel) we also plot the mass-to-number
density ratio for the sample of 299 early-type, i.e. $\eta \leq -1$
\citep{mad02}, cluster members brighter than $M_{b_J}=-19.3+5 \log h$.
The mass-to-number density profile is somewhat flatter, as expected,
since the early-type cluster galaxies have the steepest number density
profile among cluster galaxy populations \citep[e.g.][]{dre80,biv02}.
In this case, the mass-to-number density profile is consistent with
being constant, not only within the central $r < r_u$ region (0.2\%
c.l.), but also up to $2 r_u$ (74.2\% c.l.).

\section{DISCUSSION}\label{disc}
In our analysis we have determined the mass and mass-to-number density
profiles of galaxy clusters out to $\sim 2 r_{200}$. We find that both
cuspy profiles and profiles with a core are acceptable. Best-fits to
the inner $\mincir r_{200}$ region are found for a $\beta$-profile
(see eq.~\ref{eq2}) with $\beta=0.8$ and $r_c \simeq 0.06 \, r_{200}$,
or a $\xi$-profile (see eq.~\ref{eq1}) with $\xi=1.4$ and $a \simeq
0.33 \, r_{200}$. Acceptable fits are also found for a \citet{kin62}
profile with $r_c \simeq 0.14 \, r_{200}$ and a \citetalias{nfw97}
profile with $a \simeq 0.18 \, r_{200}$. Density profiles with very
large core-radii ($r_c \magcir 0.3 \, r_{200}$) are ruled out (90\%
c.l.). The density profiles that provide the best-fits to the data in
the inner $\mincir r_{200}$ region, also provide a reasonable
description of the cluster gravitational potential at larger radii,
out to $\sim 2 r_{200}$.

Let us compare our results with previous mass-profile
determinations. From the analysis of the CNOC cluster survey data-set
\citet{car97a} and \citetalias{vdm00} conclude that the average mass
profile of galaxy clusters is well described by the
\citetalias{nfw97}'s analytic form, with a scale $a \simeq 0.25 \, r_{200}$,
although other profiles are equally acceptable. Their best-fit
\citetalias{nfw97} scale is slightly larger than ours, but not
significantly so. Note anyway that the CNOC clusters are on average
more massive than the 43 clusters in our sample, so they are expected
to be less concentrated and have a larger \citetalias{nfw97} scale.

\citet{gel99} find that the \citetalias{nfw97} profile provides a good
fit to the Coma mass profile, while a softened isothermal sphere does
not. Similarly, \citet{rin00} are able to rule out a singular
isothermal sphere for the mass profile of the cluster A576, while a
\citetalias{nfw97} profile is a good fit to the observed mass
profile. Both the results of \citet{gel99} and those of \citet{rin00}
are based on the caustic analysis.

Similar conclusions about the cluster mass profile are reached by
\citet{mar99} using X-ray data. Both the cluster A2199 and the cluster
A496 have mass profiles remarkably well approximated by
\citetalias{nfw97} models, and deviate from the isothermal mass
profile (which is too steep). On the other hand, \citet*{ett02} find
that about half of 22 nearby clusters of galaxies have an X-ray
determined mass density profile which is better fitted by a
\citet{kin62}'s rather than a \citetalias{nfw97} model. In their X-ray
analyses of 12 galaxy clusters, \citet{dur94} find that the King
profile provides a good fit to the mass density distribution, but with
very small core-radii ($\simeq 15$ \kpc).

Useful constraints are derived from the gravitational lensing
analysis. Most authors \citep{clo00,clo01,lom00,bau02} conclude
in favor of a \citetalias{nfw97} mass profile and against
an isothermal sphere mass distribution, except perhaps 
\citet{she01}.

Our results are in agreement with the claims in support of a
\citetalias{nfw97} mass density profile and against both a
softened\footnote{\citet{gel99}'s softened isothermal sphere mass
profile corresponds to our $\beta$-profile with $\beta=2/3$.}  and a
singular isothermal sphere mass profiles, which both increase faster
with radius than our observed cluster mass profile. We stress however
that our data do not rule out the presence of a core in galaxy
clusters, although this core has to be very small, of the order of the
size of a galaxy.  Interestingly, our formal best-fit for a
\citetalias{nfw97} profile is obtained for $a \simeq 0.18 \, r_{200}$,
which is mid-way between the formal best-fit obtained for the CNOC
clusters by \citetalias{vdm00} and the formal best-fit obtained for
the poor galaxy systems by \citet{mah99}. Since our clusters are of
intermediate mass between the CNOC clusters and the poor systems of
galaxies of \citet{mah99}, a scaling of $a$ with the mass of the
system is suggested, as expected from numerical simulations
\citepalias{nfw97}.

Our most intriguing result is the significant decrease of the
mass-to-number density ratio at radii greater than $\sim 0.3 r_{200}$.
The mass per galaxy at $r_{200} < r < 2 r_{200}$ is only about half the mass
per galaxy in the cluster central regions. Although we do not derive
the mass-to-light profile (which would require an analysis of the
cluster luminosity function and it is beyond the scope of this paper),
this is unlikely to be constant. In fact, for the mass-to-light
profile to be flat, a strong luminosity segregation over the entire
radial range out to $\sim 2 r_{200}$ is required. Apart from the
segregation of the very bright galaxies, which only concerns the
cluster cores \citep[e.g.][]{biv02}, luminosity segregation in
clusters is also observed as a change of the number ratio of dwarf
to giant galaxies with radius\citep[e.g.][]{bei02,dur02,lob97}. However, the
luminosity segregation of dwarf galaxies is not relevant in our case,
since our number density profile is determined only on galaxies
brighter than $M_{b_J}=-19.3 + 5 \log h$, which is only $\sim 0.3$
magnitudes fainter than $M^{\star}$ in the Schechter luminosity
function for early-type galaxies \citep{mad02}. We do not find
evidence of luminosity segregation in our sample of members brighter
than $M_{b_J}=-19.3 + 5 \log h$, as their absolute magnitudes and
clustercentric distances are uncorrelated (52\% c.l., according to
a Spearman rank correlation test).

Our mass-to-number density profile implies a mass distribution more
concentrated than the galaxies. Since the hot X-ray emitting gas is
less concentrated than the galaxies \citep[e.g.][]{eyl91,dur94,cir97},
our result also implies that the {\em dark} matter distribution is
more concentrated than the {\em baryonic} matter distribution.

Previous determinations of the mass-to-light (or mass-to-number)
profiles have provided evidences both in favor
\citep{car97a,cir97,mah99,vdm00,rin01} and against
\citep{kor98,rin00,car01} a constant mass-to-light
profile. \citet{kor98} find an increasing mass-to-light profile in the
poor cluster AWM7, from the center to $\sim 0.2$ \hh. Such an increase
is also seen in our data (see Figure~\ref{f-dmdn}) but it is not
significant. \citet{rin00} find that the mass-to-light profile
decreases with clustercentric radius, in consistency with our
finding. \citet{car01} find an increasing mass-to-light profile in
galaxy groups.  Clearly, a general consensus on the relative
distribution of dark and baryonic matter in galaxy systems is still
lacking. Part of the discrepancy among different authors can arise
from the use of different photometric bands \citep{rin01}.  Cluster
members are redder in the cluster center, so that blue-band selected
cluster samples (like ours) should be characterized by flatter number
density profiles, and hence {\em steeper} mass-to-number density profiles,
as compared to samples selected in the near-infrared.  Our
mass-to-number density profile indeed becomes flatter, and consistent with
being constant, when only early-type (and hence redder) cluster
members are considered (see Fig.~\ref{f-dmdn}).

Does our result invalidates the use of simple mass estimators based on
the {\em light traces mass} hypothesis? Probably not. In fact, our
mass-to-number density profile does not significantly deviate from a
constant when we restrict the analysis to the virialized region ($r
\leq r_{200}$), or when only the early-type galaxies are considered.

\section{SUMMARY}\label{summ}
We use the June~2001 public release of the 2dFGRS to construct a
sample of 4947 galaxies with redshifts in the region of 43 clusters,
reaching out to $\sim 2 r_{200}$.  1345 cluster members are selected
with the method of \citet{fad96}. We build an ensemble cluster
profile from this sample, using $r_{200}$ and $\sigma_p$ of each
cluster to normalize the clustercentric radii and velocities of all
galaxies.

We determine the mass profile of this ensemble cluster by two
independent methods, namely those of \citetalias{vdm00} and
\citetalias{dia99}. The method of \citetalias{vdm00} is based on the
solution of the Jeans equation, and can only be applied in conditions
of dynamical equilibrium. We therefore apply it on the subsample of
cluster members within the virial radius, $r_{200}$. We exclude
late-type spirals from this sample since these galaxies have been
shown in the literature to be characterized by anisotropic orbits
\citep[e.g.][]{moh96,biv97}. The method of \citetalias{dia99} can
usefully be applied to the whole sample of galaxies in the cluster
region, since it does not require dynamical equilibrium. We use it to
constrain the cluster mass profile beyond the virial radius, and out
to the limit we imposed on our sample, $\sim 2 r_{200}$.

We find very good agreement between the mass profiles determined using
the two methods. The mass profile from the center to $\sim 2 r_{200}$
is remarkably well fitted by a $\xi$-model (eq.~\ref{eq1}) with $a
\simeq 0.33 \, r_{200}$ and $\xi=1.4$, but many other models provide
acceptable fits, including \citetalias{nfw97} models and models with a
small core ($r_c \mincir 0.2 r_{200}$). We find an uncertainty on the
total mass of $\approx \pm 8$\% at the virial radius, and $\approx \pm
15$\% at two virial radii.

We determine the mass-to-number density profile of the ensemble
cluster.  It is fully consistent with a constant from the center to
$\sim r_{200}$, but it significantly decreases at larger radii. The
decreasing trend is however not significant when the number density
profile is computed on the early-type galaxies only.

\acknowledgments 
This work is based on the 100k Data Release of the
{\em Two Degree Field Galaxy Redshift Survey,} kindly made available
to the astronomical community by the 2dFGRS Team. We acknowledge
useful discussion with Antonaldo Diaferio, Peter Katgert and Alain
Mazure. We thank an anonymous referee for her/his constructive
comments. We thank Tim Beers and David Merritt for providing us with
copies of their FORTRAN codes ROSTAT and, respectively, MAPELN. This
work was partially supported by the Italian Space Agency (ASI), and by
the Italian Ministry of Education, University, and Research (MIUR
grant COFIN2001028932 "Clusters and groups of galaxies, the interplay
of dark and baryonic matter").

\begin{deluxetable}{lccrrrr}
\tabletypesize{\small}
\tablecaption{The sample of 43 clusters}
\tablecolumns{7}
\tablehead{
\colhead{Name} & \colhead{RA$_{J2000}$} & \colhead{DEC$_{J2000}$} &
\colhead{N$_m$} &
\colhead{N$_m$}  & \colhead{$\overline{v}$} &
\colhead{$\sigma_p$} \\
\colhead{} & \colhead{($^h$ $^m$ $^s$)} & \colhead{($^{\circ}$ $'$ $''$)} 
& \colhead{} & \colhead{$(R \leq r_u)$} &
\colhead{(km~s$^{-1}$)} & \colhead{(km~s$^{-1}$)}
}
\startdata
ACO  419  & 03 06 15.3 & -23 55 57 &  41 &  30 & 20499 &   681 \\
ACO  892  & 09 51 03.1 & ~00 49 43 &  18 &  13 & 28130 &   435 \\
ACO  957  & 10 11 13.7 & -00 40 32 &  68 &  43 & 13648 &   653 \\
ACO  978  & 10 17 58.5 & -06 15 41 &  31 &  22 & 16371 &   534 \\
ACO  993  & 10 19 34.5 & -04 38 46 &  62 &  33 & 16368 &   503 \\
ACO 1098  & 10 45 26.8 & -03 40 39 &  25 &  14 & 20699 &   465 \\
ACO 1308  & 11 30 30.2 & -03 43 53 &  20 &  10 & 15496 &   287 \\
ACO 1364  & 11 40 56.5 & -01 27 40 &  37 &  24 & 31681 &   542 \\
ACO 1373  & 11 42 55.9 & -02 10 23 &  28 &  20 & 38892 &   568 \\
ACO 1750  & 13 28 45.0 & -01 28 46 &  37 &  24 & 25187 &   649 \\
ACO 2660  & 23 43 08.8 & -26 06 42 &  38 &  12 & 15968 &   825 \\
ACO 2726  & 00 04 49.4 & -28 21 52 &  31 &  15 & 18220 &   339 \\
ACO 2734  & 00 08 47.2 & -29 05 04 &  67 &  43 & 18437 &   625 \\
ACO 2915  & 01 26 25.6 & -29 17 17 &  25 &  16 & 25877 &   546 \\
ACO 2967  & 02 00 30.9 & -28 30 50 &  14 &  11 & 33036 &   498 \\
ACO 2981  & 02 07 38.0 & -27 37 23 &  14 &  10 & 32484 &   382 \\
ACO 3042  & 02 41 05.0 & -27 06 59 &  17 &  12 & 31161 &   404 \\
ACO 3814  & 21 46 25.2 & -30 58 31 &  30 &  24 & 35819 &   836 \\
ACO 3837  & 22 06 21.5 & -27 33 60 &  25 &  16 & 27390 &   366 \\
ACO 4044  & 23 46 53.2 & -27 15 59 &  22 &  15 & 32917 &   417 \\
ACO 4049  & 23 50 03.8 & -28 56 56 &  14 &   5 & 17743 &   418 \\
ACO 4053  & 23 52 10.7 & -27 57 05 &  45 &  34 & 21451 &   488 \\
ACO S246  & 02 17 39.8 & -27 26 57 &  15 &   6 & 17776 &   181 \\
ACO S301  & 02 47 28.9 & -31 21 39 & 100 &  52 &  6772 &   464 \\
ACO S340  & 03 18 04.4 & -27 12 06 &  22 &  14 & 20840 &   328 \\
ACO S983  & 21 57 12.4 & -19 23 20 &   9 &   5 & 17246 &   130 \\
ACO S1127 & 23 25 38.4 & -29 24 33 &  38 &  20 & 31259 &   767 \\
APM  171  & 01 22 38.9 & -29 45 42 &  22 &   7 & 29038 &   449 \\
APM  294  & 02 45 06.8 & -27 45 38 &  26 &  13 & 39891 &   562 \\
APM  311  & 02 52 12.3 & -33 40 18 &  27 &  20 & 32608 &   467 \\
APM  416  & 03 32 28.6 & -29 07 35 &  38 &  14 & 31360 &   656 \\
APM  715  & 21 45 56.5 & -27 53 34 &  21 &  12 & 21726 &   625 \\
APM  880  & 23 15 54.9 & -27 25 06 &  27 &   9 & 25126 &   351 \\
EDCC  69  & 21 56 06.8 & -28 42 06 &  24 &  16 &  6385 &   275 \\
EDCC 129  & 22 16 11.4 & -24 36 22 &  19 &  12 & 11400 &   397 \\
EDCC 148  & 22 25 39.8 & -24 41 59 &  36 &  16 & 23395 &   539 \\
EDCC 153  & 22 29 38.2 & -31 26 48 &  23 &  16 & 17130 &   520 \\
EDCC 155  & 22 29 20.3 & -25 39 05 &  44 &  23 & 10118 &   520 \\
EDCC 445  & 00 25 58.7 & -27 46 02 &  48 &  16 & 18511 &   483 \\
EDCC 601  & 01 51 25.0 & -33 23 52 &  23 &  16 & 29200 &   468 \\
EDCC 623  & 02 04 22.6 & -28 37 55 &  21 &  13 & 25369 &   717 \\
EDCC 641  & 02 13 53.6 & -27 47 19 &  25 &  15 & 31236 &   340 \\
EDCC 642  & 02 14 16.4 & -29 13 29 &  28 &  12 & 32513 &   375 \\
\enddata 
\label{tabsample}
\end{deluxetable} 
\end{document}